\documentclass[seceq]{ptptex}

\usepackage{graphicx}

\makeatletter
\def\simleq{\mathrel{\mathpalette\gl@align<}}
\def\simgeq{\mathrel{\mathpalette\gl@align>}}
\def\gl@align#1#2{\lower.6ex\vbox{\baselineskip\z@skip\lineskip\z@
     \ialign{$\m@th#1\hfill##\hfil$\crcr#2\crcr\sim\crcr}}}
\makeatother

\newcommand{\bra}{\langle}
\newcommand{\ket}{\rangle}
\newcommand{\braket}[1]{\bra #1 \ket}
\newcommand{\qq}{\braket{\bar{q}q}}

\newcommand{\zr}[1]{\mbox{\hspace*{#1em}}}
\newcommand{\ZZ}{\mbox{\sf Z\zr{-0.45}Z}}

\title{
Theoretical Status of Pentaquarks%
}

\author{
Takumi \textsc{Doi}$^{1,2,}$\footnote{
e-mail address: doi@pa.uky.edu
}
}

\inst{
$^{1}$ Dept. of Physics and Astronomy,
Univ. of Kentucky, Lexington, KY 40506, USA
$^{2}$ RIKEN BNL Research Center, 
BNL,
Upton, NY 11973, USA
}

\abst{
\vspace*{-5mm}
We review the current status of
the 
theoretical pentaquark search
from the direct QCD calculation.
The works from the QCD sum rule 
and the lattice QCD
in the literature are carefully examined.
The importance of the framework 
which can distinguish 
the exotic pentaquark state (if any)
from 
the NK scattering state
is emphasized.
}

\begin{document}

\maketitle

\vspace*{-9mm}
\section{Introduction}
\label{sec:intro}
\vspace*{-3mm}

While 
QCD
was established 
as a fundamental theory of the strong interaction a few decades ago,
its realization in hadron physics has not been understood
completely. For instance, 
(apparent) absence of 
``exotic'' 
states, which are different
from ordinary $q\bar{q}$ mesons and $qqq$ baryons,
has been a long standing problem.
Therefore, the announcement\cite{nakano} of the 
discovery of $\Theta^+$ (1540), 
whose minimal 
configuration is $uudd\bar{s}$, was quite striking.
For the current experimental status, 
we refer to Ref.\cite{Nakano-ykis}.

In this report, we review the theoretical effort
to search the $\Theta^+$ pentaquark state.
The main issue here is whether QCD favors
its existence or not, and the determination of possible 
quantum numbers 
for the 
pentaquark families (if any).
In particular, in order to understand the 
narrow width of $\Theta^+$ observed in the experiment,
it is crucial to determine the
spin and parity
directly from QCD.
%
%
%
%

For this purpose, we employ two frameworks, 
the QCD sum rule and the lattice QCD,
where 
both 
allow the nonperturbative QCD
calculation without models, and have achieved a great success 
for 
ordinary mesons/baryons.
Note, however, that neither of them is infallible,
and we consider them
as 
complementary
to each other.
For instance, the lattice simulation
cannot be performed at completely realistic setup,
i.e., there often exists the artifact stemming from 
discretization error,
finite volume, 
heavy u,d-quark masses
and neglection of dynamical quark effect (quenching), etc.
On the other hand, 
the sum rule can be constructed at realistic situation,
and 
is free from such artifacts 
in lattice.
Unfortunately, it 
suffer from another type of artifact.
Because a sum rule yields only the dispersion integral of 
spectrum,
an interpretive model function have to be assumed
phenomenologically.
%
%
Compared to the ordinary hadron analyses,
this procedure may weaken the predictability for the 
experimentally uncertain system, such as pentaquarks.
%
%
%
Another artifact in the sum rule is the OPE truncation:
one have to evaluate whether 
the OPE convergence is enough or not.
%
%
%
%
%
We also comment on the important issue
common to both of the methods.
Recall that the decay channel 
$\Theta^+ \rightarrow N+K$ is open experimentally.
Considering also that 
both methods calculate a two-point 
correlator
and seek for a pentaquark signal in it,
it is essential to develop a framework which
can distinguish the 
pentaquark
from the NK state in the correlator.
%
%
In the subsequent sections, we examine 
the literatures and 
see how the above-described issues 
have been resolved 
or
remain unresolved.


\vspace*{-2mm}
\section{The QCD Sum Rule Work}
\label{sec:qsr}
\vspace*{-2mm}


More than ten sum rule analyses for $\Theta^+$ spectroscopy
exist for $J=1/2$
\cite{Zhu,matheus,SDO,kek-penta,Lee1,Lee2,eidemuller,higher-dim,HJ.Lee,kojo}.
The first parity projected sum rule
was studied by us\cite{SDO} for $I=0$.
The positivity of the pole residue
in the spectral function is proposed as a
signature of the pentaquark signal.
This is superior criterion to the 
consistency check of 
predicted/experimental masses,
because it is difficult to achieve the mass prediction
within 100MeV ($\sim [m(\Theta^+) - m(NK)]$) accuracy.
We also propose the diquark exotic current
$
J_{5q} =\epsilon^{abc}\epsilon^{def}\epsilon^{cfg}
   (u_a^T C d_b) (u_d^T C \gamma_5 d_e) C\bar{s}_g^T , 
$
in order to suppress the 
NK state contamination.
The OPE is calculated up to dimension 6, checking
that the highest dimensional contribution is reasonably small.
We obtain a possible signal in negative parity.

%
%
%


In the treatment of the NK state,
improvement is proposed in Ref.\cite{Lee1}.
There, NK contamination is evaluated 
using the soft-Kaon theorem. 
Note here that the NK contamination calculated by
two-hadron reducible (2HR) diagrams
in the OPE level\cite{kek-penta}
is invalid because 
what have to be calculated is the 2HR part in the hadronic level,
not in the 
QCD (OPE)
level. 
The reanalysis\cite{Lee1} of sum rule 
up to dimension 6 shows that
the subtraction of the NK state 
does not change the result of Ref.\cite{SDO}.


Yet, as described in Sec.\ref{sec:intro},
the above sum rules 
may suffer from the OPE truncation artifact.
In fact, 
the explicit calculation
up to higher dimension
have shown that this is indeed the case\cite{higher-dim,HJ.Lee,kojo}.
Here, 
we refer to the elaborated work 
in Ref.\cite{kojo}.
They calculate the OPE for $I(J^P)=0(1/2^\pm)$
up to dimension $D=15$.
It is shown that 
the terms with $D>6$ are important as well,
while further high dimensional terms
$D > 15$ 
are not significant.
%
Another idea 
in Ref.\cite{kojo} is
the use of the combination of two independent pentaquark sum rules.
In fact, the proper combination is found to 
suppress the 
continuum contamination drastically,
which corresponds to reducing the uncertainty in
the phenomenological model function.
Examining the positivity of 
the pole residue,
they conclude the pentaquark exists
in positive parity. 

Does the result\cite{kojo} definitely predict the $J^P=1/2^+$ pentaquark ?
At this moment, we conservatively point out remaining issues.
The first problem is still the NK contamination.
While such contamination is expected to be partly suppressed
through the continuum suppression,
it is possible that 
the obtained signal corresponds to just scattering states.
In this point, Ref.\cite{kojo} argues that the signal
has different dependence on the parameter $\qq$
from the NK state.
We, however, consider 
this discussion uncertain,
because $\qq$ is not a free parameter
independent of other condensates.
For further study, the explicit estimate in the soft-Kaon limit\cite{Lee1}
is interesting check, but the calculation up to high dimension
has not been worked out yet.
%
%
Second issue is related to the OPE.
In the evaluation of the high dimensional condensates,
one have to rely on
the vacuum saturation approximation practically,
while
the uncertainty originating from this 
procedure
is not known.
%
%
Furthermore, there exists an issue
for the validity of the OPE itself when considering the sum rule
with high dimensionality.
In fact, 
rough 
analysis of the gluonic condensates 
shows\cite{SVZ} that the {\it nonperturbative} OPE may break down 
around $D \simgeq 11-16$.
One may have to consider this effect as well,
through, for instance, the instanton picture\cite{HJ.Lee}.



So far, we have reviewed $J=1/2$ sum rules.
While there are $J=3/2$ works\cite{kek-3/2,zhu-3/2},
it is likely that
they suffer from slow OPE convergence. 
Further progress is awaited.

\vspace*{-2mm}
\section{The Lattice QCD Work}
\label{sec:lat}
\vspace*{-2mm}

There are a dozen of quenched lattice 
calculations\cite{csikor12,chiu,kentucky,ishii12,lasscock12,
csikor122,alexandrou12,rabbit,holland,lasscock32,ishii32,negele,hagen}:
some of them\cite{csikor12,chiu,rabbit,lasscock32} 
report 
the positive signal,
while 
others\cite{kentucky,ishii12,lasscock12,csikor122,ishii32,holland} 
report null results.
This apparent inconsistency, however, can be understood
in a unified way, by taking a closer look at 
the ``interpretation'' of the numerical results
and the pending lattice artifact.

As discussed in Sec.\ref{sec:intro},
the question is how to identify the pentaquark signal
in the correlator, because the correlator 
at large Euclidian time 
is dominated by the ground state, 
the NK scattering state.
%
%
%
In this point, we develop a new method in Ref.\cite{ishii12,ishii32}.
Intuitively, this method makes use of that 
a scattering state is sensitive 
to the spacial boundary condition (BC),
while a compact one-particle state is expected to be insensitive.
Practically, we calculate the correlator under
two spacial BCs: 
(1) 
periodic BC (PBC) for all  u,d,s-quarks,
(2) hybrid BC (HBC) where anti-periodic BC for u,d-quarks
and periodic BC for s-quark.
The consequences are as follows.
In PBC, 
all of $\Theta^+$, N, K
are subject to periodic BC.
In HBC, while 
$\Theta^+(uudd\bar{s})$ remains subject to periodic BC,
N($uud$,$udd$) and K($\bar{s}d$,$\bar{s}u$) are
subject to anti-periodic BC.
Therefore, the energy of NK will shift 
by PBC $\rightarrow$ HBC due to the momentum of N and K,
while there is no energy shift for $\Theta^+$.
(Recall that the momentum is quantized on lattice 
as $2\vec{n}\pi/L$ for periodic BC
and $(2\vec{n}+1)\pi/L$ for anti-periodic BC,
with spatial lattice extent $L$ and $\vec{n} \in \ZZ^3$.)
In this way, the different behavior between NK and $\Theta^+$
can be used to identify whether the signal 
is NK or $\Theta^+$.
%
%
%
%
%
We simulate the anisotropic lattice, 
$\beta=5.75$, $V=12^3\times 96$, $a_\sigma/a_\tau=4$,
with the clover fermion.
The conclusion is:
(1) the signal in $1/2^-$ is found to be s-wave NK
from HBC analysis. No pentaquark is found up to $\sim$ 200MeV
above the NK threshold.
%
(2) the $1/2^+$ state is too massive ($>$ 2GeV)
to be identified as $\Theta^+(1540)$.

In comparison with other lattice results,
we introduce another powerful method\cite{kentucky} 
to distinguish $\Theta^+$ from NK.
This method makes use of that the volume dependence 
of the spectral weight
behaves as ${\cal O}(1)$ for one-particle state,
and as ${\cal O}(1/L^3)$ for two-particle state.
Intuitively, the latter factor ${\cal O}(1/L^3)$
can be understood as the encounter probability of the two particles.
The calculation\cite{kentucky} of the spectral weight from
$16^3\times 28$ and $12^3 \times 28$ lattices 
reveals that 
the ground states of both the $1/2^\pm$ channels
are not the pentaquark, but the scattering states.
Further analysis is performed in Ref.\cite{rabbit}.
There, the 1st excited state in $1/2^-$ 
is extracted with $2\times 2$ variational method.
The volume dependence of the spectral weight 
indicates that the 1st excited state
is not a scattering state
but a pentaquark state.
This is consistent with Ref.\cite{negele},
where $19\times 19$ variational method is used to extract the excited states.

Note here that this results is {\it consistent}
with the HBC analysis\cite{ishii12}.
In fact, HBC analysis exclude the pentaquark up to $\sim$ 200MeV above
threshold, while the resonance observed in Ref.\cite{rabbit} locates
200-300MeV above the threshold.
The question is whether the observed resonance is really 
$\Theta^+$ which experimentally locates 100MeV above the threshold.
To address this question, explicit simulation is necessary
at physically small quark mass without quenching.
In particular, small quark mass would be important
considering that Refs.\cite{ishii12,rabbit} are simulated at rather heavy quark masses
and expected to suffer from large uncertainty in the chiral extrapolation.

Finally, we discuss the $J^P=3/2^\pm$ lattice results.
We performed the 
comprehensive study\cite{ishii32} with three different operators
and conclude that 
all the lattice signals are too massive ($>$ 2GeV) for $\Theta^+$,
and are identified as not pentaquarks but scattering states
from the HBC analysis.
On the other hand,
Ref.\cite{lasscock32} claims that
a pentaquark candidate is found in $3/2^+$.
We, however, observe that the latter result 
are contaminated by significantly large statistical noise,
which makes their result quite uncertain.
Note also that their criterion 
to distinguish $\Theta^+$ 
from scattering states
is based on rather limited 
argument
compared to the HBC analysis.

\vspace*{-2mm}
\section{Conclusions}
\label{sec:conclusion}
\vspace*{-2mm}

We have examined both of the QCD sum rule and lattice QCD works.
In the sum rule, progresses in OPE calculation and continuum suppression
have achieved stable 
analysis,
while the subtraction of NK contamination remains a critical issue.
In the lattice, the framework which distinguish 
the pentaquark from NK have been successfully established.
In order to resolve the {\it superficial} inconsistency in the
lattice prediction, 
the calculation at small quark mass without quenching
is highly desirable.


\vspace*{-2mm}
\section*{Acknowledgements}
\vspace*{-2mm}

This work is completed in collaboration with
Drs.
H.Iida, 
N.Ishii, 
Y.Nemoto, 
M.Oka, 
F.Okiharu, 
H.Suganuma
and
J.Sugiyama.
T.D. is supported 
by Special Postdoctoral Research Program of RIKEN
and by U.S. DOE grant DE-FG05-84ER40154.


%



\end{document}